\documentclass[12pt]{iopart}

  \expandafter\let\csname equation*\endcsname\relax
  \expandafter\let\csname endequation*\endcsname\relax

\usepackage{amstext,amsmath,amssymb,amsfonts,bm,bbm}
\usepackage[usenames,dvipsnames]{color}
\usepackage{graphicx}
\usepackage{pfnote}
\usepackage{euscript}
\usepackage{hyperref}

 \usepackage{color} 

\makeatletter
\@ifundefined{textcolor}{}
{%
 \definecolor{BLACK}{gray}{0}
 \definecolor{WHITE}{gray}{1}
 \definecolor{RED}{rgb}{1,0,0}
 \definecolor{GREEN}{rgb}{0,.4,0}
 \definecolor{BLUE}{rgb}{0,0,1}
 \definecolor{CYAN}{cmyk}{1,0,0,0}
 \definecolor{MAGENTA}{cmyk}{0,1,0,0}
 \definecolor{YELLOW}{cmyk}{0,0,1,0}
 }

\newcommand{\be}{\begin{equation}}
\newcommand{\ee}{\end{equation}}
\def\bal{\begin{align}}
\def\eal{\end{align}}
\def\bay{\begin{array}}
\def\ear{\end{array}}

\def\0{\otimes}
\def\1{{{\mathbbm 1}}}
\def\6{\langle}
\def\9{\rangle}

\def\half{\mbox{$1\over2$}}

\def\hp{\hat {p}}
\def\hx{\hat {x}}

\def\hH{\hat{H}}
\def\hK{\hat{K}}
\def\mhU{\hat{U}}
\def\mhV{\hat{V}}
\def\hW{\hat{W}}

\def\mR{\mathbb{R}}
\def\mZ{\mathbb{Z}}

\def\hpo{\mathcal{H}_{\mathrm{poly}}}

\def\cyl{\mathrm{Cyl}}
\def\sha{\mathrm{shad}}

\def\e0{\epsilon_0}

\def\sch{Schr\"{o}dinger~}

\def\0{\otimes}
\def\1{{{\mathbb 1}}}
\def\6{\langle}
\def\9{\rangle}

\def\half{\mbox{$1\over2$}}

\def\pad{\partial}

\def\schs{Schr\"{o}dinger~}
\def\sch{Schr\"{o}dinger}

\begin{document}

\renewcommand{\thefootnote}{\arabic{footnote}}

\title[Entropy and entanglement in polymer quantization]{Entropy and entanglement in polymer quantization}

\author{Tommaso~F Demarie$^{1,2}$ and Daniel~R Terno$^{1}$}

\address{$^1$Department of Physics and Astronomy, Faculty of Science, Macquarie University, 2109 NSW, Sydney, Australia}

\address{$^2$Centre for Engineered Quantum Systems (EQuS), Macquarie University, 2109 NSW, Sydney, Australia}

\eads{\mailto{tommaso.demarie@mq.edu.au}, \mailto{daniel.terno@mq.edu.au}}

\begin{abstract}
Polymer quantization is a useful toy model for  the mathematical aspects of loop quantum gravity and is interesting in its own right. Analyzing entropies of physically equivalent states in the standard Hilbert space and the polymer Hilbert space we show that they converge in the limit of vanishing polymer scale. We derive a general bound that relates entropies of physically equivalent states in unitarily inequivalent representations.

\end{abstract}


\date{\today}
\maketitle

\section{Introduction}

Statistical thermodynamics, quantum mechanics and information theory promoted entropy from an auxiliary variable of the mechanical theory of heat \cite{claus} to  one of the most important quantities in  science \cite{law2, wehrl, op-e}.
  Microscopic derivations of the thermodynamical entropy were, and still are, a major part of the statistical physics.
Entropy and related quantities underpin most of the classical and quantum information theory \cite{qinfo}.   Black hole thermodynamics \cite{bhte} introduced the idea that an upper bound on the entropy of a system (and entanglement between the subsystems) scales with the area, and not with the volume.  Formalized as a holographic principle, the area law was demonstrated in a variety of situations {\cite{holo,fmw,holo-g}}, which include string theory \cite{stringref} and loop quantum gravity (LQG) \cite{lqg}, calculations of black hole entropy \cite{bhte,bh-1} and related models \cite{lt}.

LQG and the spin networks/spin foams formalisms represent the idea of fundamental discreteness. They are conceptually and   technically  different  from the usual quantum field theory on a  space-time continuum. This difference leads to a number of questions. One of them \cite{lqg, ash03} is   emergence of the semiclassical states     of quantum gravity. These states have to enable the   space-time continuum view and the ensuing construction of a field theory. Investigation of this problem and the associated mathematical issues were one of the motivation to study the polymer quantization of a single non-relativistic particle  \cite{ash03} as a useful toy model.

Another question pertains to the validity of  holographic principle(s) and compatibility of the entropy calculations across different quantization schemes. Recent advances in coupling gravity and matter \cite{gs-ln} in   LQG  make it necessary to re-consider   holography in these systems,  since it  is usually derived separately  in the gravity and the matter sectors.  One can wonder if two unitarily inequivalent quantization schemes give compatible predictions for   observables, are the   predicted entropies ``close"?   While the desirable answer is ``yes", its validity is neither obvious nor guaranteed. Again, the polymer quantization provides a convenient model setting to study this problem.

We demonstrate that if the \schs and the polymer schemes  give ``close" (to be precisely defined below) predictions for fundamental observables, their predictions of entropy are also close. The two entropies coincide in the continuum limit of the polymer quantization. We illustrate this by comparing the entanglement entropy in the ground state of two coupled harmonic oscillators in the two schemes. Our results, derived from first principles, agree and justify the thermodynamic considerations of \cite{sigma}.

The rest of this paper is organized as follows. After a brief review of the polymer quantization we describe our main result, outline the calculational techniques for Gaussian states, and present the  example. We conclude by discussing  possible limitations and future directions.

\section{Review of polymer quantization}

A useful starting point  is a unitary  Weyl (displacement) operator,
\be
\hW(q,k)^\dag=\hW(-q,-k),
\ee
where $q$ and $k$ are the translation an boost parameters, respectively. The Weyl operator can be expressed as a product of the translation and boost operators,
\be
\hW(q,k)=\rme^{\frac{\rmi}{2}qk} \mhV(q)\mhU(k)=\rme^{-\frac{\rmi}{2}qk} \mhU(k)\mhV(q).
\ee
{The product of Weyl operators
\be
\hW(q_1, k_1) \hW(q_2, k_2) = e^{-\frac{i}{2} (q_1 k_2 - q_2 k_1)} \hW(q_1+q_2, k_1+k_2) \, ,
\ee
and the unitarity condition
\be
\hW(q,k)^\star = \hW(-q,-k) \, ,
\ee
define the   Weyl algebra}. Note that here and in the following we set $\hbar=1$.

In the standard \schs representation {of the Weyl algebra} {the translation and boost} operators are generated by the {canonical} self-adjoint  position and momentum operators,
\be
\mhV(q)=\rme^{- \rmi q\hp }, \qquad \mhU(k)=\rme^{\rmi k\hx},
\ee
and the Weyl operator acts on wave functions as
\be
\hW(q,k)\psi(x)= \rme^{-\frac{\rmi}{2} kq}\rme^{\rmi kx}\psi(x+q).
\ee
 {The polymer representation is an alternative irreducible representation of the Weyl algebra where the quantum states are represented by countable sums of plane waves. In this sense the polymer Hilbert space $\hpo$ is} built around (Harald) Bohr compactification of the real line \cite{halv,gam69}.
A slightly different construction is possible by noting that the (position) space $\hpo$ is spanned by the basis states $|x\9$,
\be
\psi_x(x')=(x'|x\9=\6x'|x\9=\delta_{x'x}, \label{basis}
\ee
where $\delta$ is the Kronecker's delta, and the various bracket signs  are explained below. Countable superpositions of these states,
\be
|\Psi\9=\sum_{x\in\gamma}\psi(x)|x\9,
\ee
where $\gamma$ is a countable set (a ``graph" \cite{ash03}) and functions $\psi$ satisfy certain fall-off conditions, form a space of cylindrical functions $\cyl_\gamma$. The infinite-dimensional space $\cyl$ is a space of functions that are cylindrical with respect to some graph. The Hermitian inner product on Cyl follows from \eqref{basis},
\be
\6\Psi|\Psi'\9=\sum_{x\in\gamma\cap\gamma'}\!\!{\psi}^*\!(x)\psi'(x).
\ee
The Cauchy completion of Cyl is $\hpo$, and the   triplet of spaces
\be
\cyl\subset\hpo\subset\cyl^\star,
\ee
where $\cyl^\star$ is an (algebraic) dual of Cyl, shares similarities with the construction of the physical space of LQG \cite{ash03}. The questions of topology are discussed in {\cite{ash03,fred06,cor07,velhinho}}.
Dual elements are denoted by $(\Upsilon|$, and their action on the elements of $\hpo$ by $(\Upsilon|\Psi\9\equiv\Upsilon(\Psi)$. The inner product defines a dual element by
\be
\Phi(\Psi)\equiv\6\Phi|\Psi\9.
\ee
We introduce the momentum dual states
$(p|\in\cyl^\star$ by
\be
(p|x\9\equiv \rme^{-\rmi px}.
\ee
Hence
\be
(p|=\sum_{x\in\mR} \rme^{-\rmi p x} {(x|}, \qquad \psi(p)=(p|\Psi\9.
\ee
In the ``position representation" that was described so far the operator $\mhU(k)$ has a self-adjoint generator $\hx$, that acts by multiplication and has normalized eigenstates,
\be
\hx|x\9=x|x\9.
\ee
However, since $|x_j\9$ and $\mhV(q)|x_j\9$ are orthogonal to each other no matter how small the translation is, the translation operator is not weakly continuous and the momentum operator is undefined. Similarly, while it is easy to show that the usual relation
\be
(p|\mhV(q)=\rme^{\rmi pq}(p|
\ee
holds, it does not correspond to any state of Cyl.

 On the other hand, a  momentum representation  is spanned by the states $|k_j\9$. In this case there exists a self-adjoint momentum operator with normalizable eigenstates, but the position operator is undefined. Since the polymer representations are not weakly continuous,  they do not satisfy the conditions of the Stone-von Neumann \cite{br2} theorem and thus are not unitarily equivalent to the \schs representation \cite{ash03,halv}.

 The two constructions of the $\hpo$ allow the most extreme expression \cite{halv} of Bohr's complementarity \cite{bohr_wz}. Failure of the weak continuity makes it into a convenient toy model of LQG, and it was studied both as such and in its own right \cite{ash03,fred06,cor07,app}. Moreover, the polymer quantization and its generalization to a scalar field were applied to the problems of quantum cosmology and spherically-symmetric collapse \cite{viqar}.

 Fell's theorem \cite{fell} guaranties that, although two
representations may be unitarily inequivalent, by using finite-precision  expectation values of a
finite number of  observables it is impossible to distinguish between the two. More precisely, for: A state $\rho_1$ of one of the representations, a finite set of operators $A_i$ that belong to the Weyl algebra and whose expectations are calculated on that state, and the set of tolerances $\epsilon_i$, there exists a ``physically equivalent" state $\rho_2$ on another representation  resulting in the expectation values that differ from the first set  by less than the prescribed tolerances.   Since its proof is not constructive, an explicit construction of states and operators is needed in each case.

 While a mathematically rigorous construction involves a multi-scale lattice \cite{fred06}, we follow the approach of \cite{ash03} and will work with a regular lattice $\gamma$, where the neighboring points have a fixed spacing $\mu$.
 An effective momentum operator is introduced through the finite difference
 \be
 \hp_\mu\equiv -\frac{\rmi}{2\mu}\big(\mhV(\mu)-\mhV^\dag(\mu)\big).
 \ee
The limit $\mu\rightarrow0$   in the \schs representation gives the usual momentum operator $\hp$. Similarly, its square is
\be
\hp_\mu^2\equiv(\hp_\mu)^2=\frac{1}{4\mu^2}\big(2-\mhV(2\mu)-\mhV(-2\mu)\big) \label{msq}.
\ee
The operators are self-adjoint \cite{ash03,fred06}. However, $\hp_\mu^2$ is not positive on a general $\cyl_\gamma$ even if it is a regular lattice. For example, for a state \cite{fred06}
\be
|\Psi\9=\sum_{x\in\gamma}\rme^{\rmi kx}\rme^{{x^2}\!/{2d^2}}|x\9
\ee
the expectation value $\6\Psi|\hp_\mu^2|\Psi\9$ is real only for a symmetric graph, i.e. $x\in\gamma\Leftrightarrow -x\in\gamma$. We will restrict ourselves only to such graphs.

It follows from \eqref{msq} that $\hp_\mu^2$ skips the neighboring lattice sights when acting on states, hence its eigenfunctions can have support on either even- or odd-numbered sights $n\mu$, $n\in\mZ$.  As a result, using $\hp_\mu^2$ as the kinetic terms for a Hamiltonian $\hH_\mu$ leads to a double degeneracy of the eigenstates \cite{ash03,app} when compared with the \schs representation.  Without appealing to it, one can note that the state  counting gives twice the semiclassical result $\int\! dp\,dx/2\pi$. Depending on one's goals it is possible either to adjust the state counting or to introduce a kinetic operator \cite{ash03}
\be
\hK_\mu\equiv\widehat{p^2}_{\mu}\equiv\frac{1}{\mu^2}\big(2-\mhV(\mu)-\mhV(-\mu)\big) \label{kin}.
\ee

The commutation relations between the operators are
\begin{align}
\nonumber
&[\hx,\mhV(\mu)] =-\mu\mhV(\mu),  \\ & [\hx,\hp_\mu]=\frac{\rmi}{2}\big(\mhV(\mu)+\mhV(-\mu)\big)=\rmi \big(1-\half\mu^2\hK_\mu\big),
\\ \nonumber & [\hp_\mu,\hK_\mu]=0.
\end{align}

A key step in extracting physical predictions in the polymer quantization is to consider the shadow states \cite{ash03} that realize the physically equivalent state on $\hpo$.  Given a  state $| \Psi \rangle \in \cyl_\gamma$, a shadow state corresponding to the dual element $(\Psi|$ is constructed such that its on $| \Psi \rangle$ is equivalent to the scalar product between its shadow and the same state, i.e.
\be
(\Phi| \Psi \rangle = \langle \Phi^\sha_\gamma | \Psi \rangle,
\ee
so the shadow states
are projections of the elements of $\cyl^\star$ onto $\cyl_\gamma$,
\be
(\Phi|=\sum_{x\in\mR}\!\Phi^*\!(x)(x|\rightarrow |\Phi^\sha_{ {\gamma}}\9=\sum_{x\in\gamma}\Phi(x)|x\9.
\ee
Using these states it is possible to explicitly demonstrate physical indistinguishability of the predictions of the polymer and \schs quantizations, as mandated by Fell's theorem, {in the sense that the expectation values of the observables of interest, calculated using shadow states, fall within a prescribed range of tolerance from the \schs predictions}.

\section{Entropy}
{Our goal is to find a relationship between the von Neumann entropy of a state $\rho$ (in the \schs representation),
\be
S(\rho) = - \text{tr} \rho \log{\rho},
\ee
and of it polymer analog $\rho^\mu$. There are several reasonable ways to relate the states, and} unlike the case of observables, there is no existent theorem that guaranties equivalence of the entropy predictions. Moreover, it is known \cite{wehrl} that entropy is not a continuous function, and without additional restrictions  there are states of infinite entropy in a neighborhood of any state.\\
{We review several properties of  entropy that will be used in the following (see \cite{wehrl, op-e} for more details). Consider first  a convex combinations of some states $\rho_i$,
\be
\rho = \sum_i w_i \rho_i\, , \quad \sum_i w_i = 1\, , \quad \forall \, w_i > 0\, .
\ee
The concavity of entropy
\be
S(\rho) = S \left(\sum_i w_i \rho_i \right) \ge \sum_i w_i S(\rho_i)
\ee
results in the upper bound
\be
S(\rho) \le \sum_i w_i S(\rho_i) - \sum_i w_i \log{w_i}\, ,
\ee
which becomes $S(\rho) \le - \sum_i w_i \log{w_i}$ if all the states $\rho_i$ are one-dimensional projectors.\\
We will make use of another property of entropy, the lower semicontinuity. If a sequence of density matrices $\rho_k$ weakly converges (i.e. all matrix elements satisfy $\6 l | \rho_k | l \9 \to \6 l | \rho | l \9$) to the density matrix $\rho$
\be
\rho_k \stackrel{weakly}\longrightarrow \rho \, ,
\ee
then the entropy of $\rho$ is bounded by
\be
S(\rho) \le \text{lim} \, \text{inf} \, S(\rho_k) \, .
\ee
Since $\rho$ is a density matrix, the weak convergence actually implies $\text{tr} |\rho_k - \rho| \to 0$. It can be shown that the lower semicontinuity of the relative entropy $S(\sigma| \rho)$ in $\rho$ implies the lower semicontinuity of the free energy at the temperature $\beta^{-1}$
\be
F(H,\rho,\beta):= \text{tr} \rho H - S(\rho)/\beta \, ,
\ee
where $H$ is the Hamiltonian of the system, and the state $\rho$ is not necessarily thermal.\\
This has an important consequence: if in addition to the weak convergence of the states (i.e. $\text{tr} |\rho_k - \rho| \to 0$), also the energy expectation values converge
\be
\text{tr} \rho_k H \to \text{tr} \rho H\,,
\ee
then the entropies converge as well,
\be
\label{conveq}
S(\rho) = \lim{S(\rho_k)} \, .
\ee

The relationship between the state $\rho$ and its polymer analog can be established in several ways that we now consider. The simplest case is when the \schs state
\be
\rho = \sum_i w_i | \psi_i \9 \6 \psi_i |\, , \quad \sum_i w_i = 1\, , \quad \forall \, w_i > 0
\ee
is a mixture of the eigenstates of some operator, and its polymer counterpart has the corresponding eigenbasis $| \Psi_i\9$. The decomposition
\be
\rho^{\mu} = \sum_i w_i | \Psi_i\9 \6 \Psi_i |
\ee
then trivially has the same entropy. In this case it is natural  to argue that the expectation values of all the observables of interest in two representations are close, but it should be established in the case-by-case analysis.

Consider now the analog
 \be
\rho^\mu=\sum_i w_i |\Psi(\mu)_i \9  \6\Psi(\mu)_ i|,
\ee
 where $|\Psi(\mu)_i\9$,  the normalized ``close approximations" of the states $|\psi_i\9$  in the sense of Fell's theorem, are mixed with the same weights.  We assume that the states $|\Psi(\mu)_i\9$ are pure, but not necessarily orthogonal.

Concavity of entropy \cite{wehrl} leads to an inequality that in our case reads as
\be
S(\rho_\mathrm{shad}^\mu)\leq -\sum w_i\log w_i=S(\rho). \label{conc}
\ee
This result holds for any two representations where the equivalent state of any pure $|\psi\9$ is pure.

A closely related result is derived as follows. In the following we will use the harmonic oscillator eigenstates both in the \schs and polymer representations, labeling the states as $|n\9$ and $|n^\mu \9$, respectively. If the \schs state is projected on the regular lattice with spacing $\mu$, resulting in $| n^\mu_{\mathrm{shad}}\9$, then \cite{ash03}
\be
| n^\mu \9 = | n^\mu_{\mathrm{shad}}\9 + | \Delta n^\mu\9\, , \qquad  \6 \Delta n^\mu| \Delta n^\mu\9 \overrightarrow{ _{\mu\rightarrow 0}}\, 0\, . \label{apshad}
\ee
In these bases the two physically equivalent states $\rho$ and $\rho^\mu$ can be written as
\be
\rho=\sum_{kl} w_{kl}|k\9\6l|\, \Longleftrightarrow \, \rho^\mu = \sum_{kl} (w_{kl} + \Delta w_{kl}^\mu) |k^\mu \9 \6 l^\mu | \, ,
\ee
for some $\Delta w_{kl}^\mu$. This quantity converges to zero as to ensure the agreement for the probabilities for projecting on the states $\alpha |m\9 + \beta|n\9$ and their polymer analogs.\\
Consider now a classical Hamiltonian of the form
\be
H = \frac{1}{2m} p^2 + V(x) \, .
\ee
We can establish the separate convergence of expectations of the kinetic and potential energy terms as follows. The kinetic term is given by Eq.(\ref{kin}), and is built from  the difference  of two Weyl algebra operators. Hence the direct use of the Fell's theorem guaranties that
\be
\lim_{\mu \to 0} \text{tr} \rho^\mu \hat{K}_\mu = \text{tr} \rho \hat{p}^2  \, .
\ee
On the other hand, using Eq.~(\ref{apshad}) for all potentials $V(x)$ such that $\6 l| V| m\9$ is finite, the projection onto the lattice and the subsequent summation form the Riemannian sum for the above \schs integral expression. For $V(x)$ growing sub-exponentially with $x$ we have $\6 \Delta l^\mu | V | m^\mu_{\mathrm{shad}} \9 \to 0$ with $\mu \to 0$, hence
\be
\6 l^\mu | V(\hat{x}) | m^\mu \9 \to \6 l^\mu_{\mathrm{shad}} | V | m^\mu_{\mathrm{shad}} \9 \to \6 l| V(\hat{x}) | m \9 \, ,
\ee
where the potential in the first term refers to polymer quantization and in the last term to \schs quantization.

Since both the matrix elements of the density matrices and the energy expectation values converge, Eq.(\ref{conveq}) applies and we can establish the desired convergence
\be
\lim_{\mu \to 0}S(\rho^\mu) = S(\rho) \, .
\ee

\section{Coupled harmonic oscillators}

We illustrate our result by considering entanglement of the ground state of two position-coupled oscillators with the quadratic Hamiltonian
\be
H = \frac{1}{2 m} (p_1^2 + p_2^2) + \frac{m \omega^2}{2} (x_1^2+x_2^2) + \lambda(x_1 - x_2)^2\, ,
\ee
 where $\lambda < m \omega^2/2$ is a positive coupling constant.

The parameter that determines the closeness of the physical predictions is the ratio of the lattice size $\mu$ to the oscillator scale
\be
d=(m\omega)^{-1/2}.
\ee
The ground state of this Hamiltonian (in the \schs quantization) is Gaussian \cite{gauss,tom}, i.e., all statistical moments of the canonical observables can be expressed from the first two --- the expectation values and the variances. Ground states of any quadratic $n$-particle Hamiltonian, as well as thermal states, coherent states and squeezed states are of this type. Gaussian states are very important in quantum optics and quantum information processing. They  have a number of useful mathematical properties, of which we need two. First, a reduced density matrix of  a Gaussian state  (e.g.,   density operator of the first oscillator in the above example) is also Gaussian. Second, entropy of a Gaussian state can be calculated using its symplectic eigenvalues, as we now describe \cite{holo-g,gauss,tom}.

Position and momenta (either classical or quantum) of $n$ particles can be arranged into a single 2$n$-dimensional vector $\mathbf{r}^T=(x_1,\ldots, x_n,p_1,\ldots, p_n)$. Classical Poisson brackets and quantum commutation relations can be expressed using the $2n\times 2n$ symplectic matrix $\mathbf{J}$,
\be
\mathbf{J} = \left(\begin{tabular}{cc}
0 & $\mathbf{\mathbb{I}}_n$ \\
$-\mathbf{\mathbb{I}}_n$ & 0
\end{tabular}\right), \qquad [\hat{\mathbf{r}}_k,\hat{\mathbf{r}}_{l}]=\rmi \mathbf{J}_{kl},
\ee
where $\mathbf{\mathbb{I}}_n$ is the $n$-dimensional identity matrix, and the symplectic matrix satisfies $\mathbf{J}^T=-\mathbf{J}=\mathbf{J}^{-1}$. The Hamilton equations are given by $\dot{r}=J\pad H/\pad r$, and the canonical transformations are represented by matrices $\mathbf{Y}$ that satisfy $\mathbf{Y}\mathbf{J}\mathbf{Y}^T=\mathbf{J}$, thus forming the $n$-dimensional symplectic group.

The vector of expectation values   and the symmetric correlation matrix are defined by
\be
\mathbf{D}=\6\hat{\mathbf{r}}\9, \qquad \mathbf{\Gamma}_{ij}=\big\langle\{(\hat{\mathbf{r}}_i - \mathbf{D}_i  ),(\hat{\mathbf{r}}_j - \mathbf{D}_j )\}\big\rangle
\ee
where $\{a,b\}$ is the anticommutator. Unitary operators $\hat{U}(\mathbf{Y})$ that, by transforming the Gaussian state according to $\rho'=\hat{U}(\mathbf{Y})\rho \hat{U}^\dag(\mathbf{Y})$, transform the statistical moments according to
\be
\mathbf{D}'=\mathbf{Y}\mathbf{D}, \qquad \mathbf{\Gamma}'=\mathbf{Y}\mathbf{\Gamma} \mathbf{Y}^T,
\ee
are called Gaussian operators. They represent important experimental procedures in quantum optics and quantum information \cite{gqi}.

Any Gaussian correlation matrix $\mathbf{\Gamma}$ can be diagonalized by some symplectic transformation $\mathbf{Y}_W$, {such that}} $\mathbf{Y}_W\mathbf{\Gamma} \mathbf{Y}_W^T=\mathbf{W}$, where $\mathbf{W}$ is a diagonal matrix $\mathrm{diag}(\sigma_1,\ldots,\sigma_n,\sigma_1,\ldots,\sigma_n)$. This spectrum consists of  the   positive values of the $n$ eigenvalue pairs   of the matrix product $\rmi \mathbf{J} \mathbf{\Gamma}$ \cite{tom}.  The eigenvalues $\sigma_j$ are   called the symplectic eigenvalues   of the covariance matrix. Finally, entropy of a $n$-particle Gaussian state $\rho$ is given by its symplectic eigenvalues as  {\cite{holo-g}}
\be
S(\rho)=\sum_{j=1}^n\left(\frac{\sigma_j+1}{2}\log_2\frac{\sigma_j+1}{2}-\frac{\sigma_j-1}{2}\log_2\frac{\sigma_j-1}{2}\right). \label{ent-cor}
\ee

Covariance matrix of the ground state of a single  harmonic oscillator (in the \schs representation) is
\bal
\mathbf{\Gamma} = \left (
\begin{tabular}{c c}
$2 \langle \hx^2 \rangle$ & $\langle \hx \hp \rangle + \langle \hp \hx \rangle$\\
$\langle \hp \hx \rangle + \langle \hx \hp \rangle$ & $2 \langle  \hp^2 \rangle$\\
\end{tabular} \right ) =
  \left (
\begin{tabular}{c c}
$d^2$ & $0$\\
$0$ & $ d^{-2}$\\
\end{tabular} \right ) \, .
\end{align}

Calculations in the polymer representation also give zero expectations for the ground state of the harmonic oscillator, $\6\hx\9=0$, $\6\hp_\mu\9=0$, but the variances are \cite{ash03}
\be
\6\hx^2\9\approx \frac{d^2}{2}\left(1-\frac{4\pi^2d^2}{\mu^2}e^{-\pi^2 d^2\!/\mu^2}\right), \qquad \6\hp_\mu^2\9\approx\frac{1}{2d^2}\left(1-\frac{\mu^2}{2d^2}\right).
\ee
The correlation term vanishes exactly. Hence, keeping only the leading terms in $\mu/d$  the correlation matrix becomes
\be
\mathbf{\Gamma}_\mu=\left(
\begin{tabular}{cc}
$d^2$ & $0$ \\
$0$ & $\frac{1}{d^2}\left(1-\frac{\mu^2}{2d^2}\right).$
\end{tabular}\right)
\ee
The product of uncertainties {in} the polymer quantization is less than its standard value of $\tfrac{1}{4}$ \cite{ash03,fabio}. The state is no longer exactly Gaussian:  the correlation matrix violates the inequality $\mathbf{\Gamma}\mu+\rmi \mathbf{J}\geq 0$ \cite{gauss}, and \eqref{ent-cor} gives a complex  value of the entropy,  $S \sim \rmi \mu^2 / d^2$, for a pure state.

The standard measure of a pure state bipartite entanglement is the von Neumann entropy of either of the reduced density matrices \cite{qinfo}. For the Gaussian states it can be calculated using \eqref{ent-cor}, with symplectic eigenvalues for the correlation matrix of the subsystem. Transforming to the normal modes (this is a symplectic transformation) gives two uncoupled oscillators with frequencies
\be
\omega_1=\omega, \qquad  {\omega_2} = \omega \sqrt{1 + \frac{4 \lambda}{m \omega^2}}\equiv\omega\alpha>\omega.
\ee
The (\sch) correlation matrix in the normal coordinates is $\mathbf{\Gamma}'=\mathrm{diag}(d^2,d^2 \alpha^{-1},d^{-2},d^{-2}\alpha)$, and the only eigenvalue of the reduced correlation matrix is
\be
\sigma_1 = \frac{1 + \alpha}{2 \sqrt{\alpha}}.
\ee

The polymer quantization is treated similarly. One needs to keep in mind that it is necessary to perform the symplectic transformation before quantization, as these two procedures do not commute.   Using $\alpha>1$ and keeping only the leading order terms in the powers of $\mu/d$ we find
\be
\sigma_1^\mathrm{poly}=\sigma_1-\frac{(1+\alpha^2)}{8\sqrt{\alpha}}\frac{\mu^2}{d^2},
\ee
and
\be
S^\mathrm{poly}=S-\frac{1+\alpha^2}{8\sqrt{\alpha}} { {\left (\log{\frac{\sqrt{\alpha}+1}{\sqrt{\alpha}-1} }\right )}}\frac{\mu^2}{d^2} \, ,
\ee
 {in agreement with the results of the previous Section.

\section{Summary and outlook}
We showed that the shadow state construction gives not only ``close" expectation values for the observables in the unitarily inequivalent representations, but the entropies of the two states agree as well. The lattice effects modify the expectation values and commutation relations. Shadow states do not satisfy the Gaussian property exactly, but only up to the terms of the order of $\mu^2/d^2$.
Our results give also physical justification to the entropy corrections derived in \cite{sigma} for the statistical thermodynamic properties of one-dimensional polymer quantum systems.

Since the convergence of entropy of the shadow states to the \schs representation value is not necessarily uniform, the following scenario is plausible. Both the discrepancy in the expectation of the momentum variances and  symplectic eigenvalues are of the order of $\mu^2/d_j^2$ for each (uncoupled) oscillator. This is also order of magnitude of  the corresponding change in entropy contribution if \eqref{ent-cor} is used.  Hence, even if the expectations of the observables agree, for a fixed value of $\mu$ and a large number $n$ of oscillators the two entropies will differ by $\sim n\mu^2/d^2$, which may be a significant amount. We leave for future work a precise estimate of the discrepancy in entropies for a fixed scale and a large number of particles.

\ack
We thank M. Cirio, M. Horodecki, P. Horodecki, V. Hussain, and  R. Verch for useful discussions and J. Louko for suggestions and critical comments. T.F.D. received support from the ARC via the Centre of Excellence in Engineered Quantum Systems (EQuS), Project No. CE110001013.


\section*{References}

\begin{thebibliography}{99}
\bibitem{claus} Clausius R 1851 {\it Phil. Mag. Series} 4 \textbf{2} 1; Clausius R 1856 {\it Phil. Mag. Series} 4 \textbf{12} 81.
\bibitem{law2} Lieb E H and Yngvason Y 1999 {\it Phys. Rep.} \textbf{310} 1.
\bibitem{wehrl} Wehrl A 1978 \RMP \textbf{50} 221.
{\bibitem{op-e} Ohya O and Petz D 2004 {\it Quantum Entropy and Its Use} (Berlin: Springer, 2nd edition).}
\bibitem{qinfo} Bru{\ss} D and Leuchs G 2007 {\it Lectures on Quantum Information} (Weinheim: Wiley-VCH); Nielsen M A and Chuang I L 2000 {\it Quantum Computation and Quantum Information} (Cambridge: Cambridge University); Peres A and Terno D R 2004 \RMP \textbf{76} 93.
\bibitem{bhte} Bekenstein J D 1974 \PR D \textbf{9} 3292; Wald R M 2001 {\it Living Rev. Relativ.} \textbf{4} 6; Padmanabhan T 2010 {\it Rep. Prog. Phys.} \textbf{73} 046901.
\bibitem{holo} Bousso R 2002 \RMP \textbf{74} 825.
{\bibitem{fmw} Flanagan E E, Marolf D and Wald R 2000 \PR D \textbf{62} 084035; Bousso R, Flanagan E E and Marolf D 2003 \PR D \textbf{68} 064001.}
\bibitem{holo-g} Eisert J, Cramer M and Plenio M B 2010 \RMP \textbf{82} 277.
\bibitem{stringref} Maldacena J M, 1998 {it Adv. Theor. Math. Phys.} \textbf{2} 231.
\bibitem{lqg} Rovelli C 2004 {\it Quantum Gravity} (Cambridge: Cambridge University Press); Thiemann T 2007 {\it Modern Canonical Quantum General Relativity} (Cambridge: Cambridge University Press).
\bibitem{bh-1} Rovelli C 1996 \PRL \textbf{77} 3288; Ashtekar A, Baez J and Krasnov K 2000 {\it Adv. Theor. Math. Phys.} \textbf{4} 1; Domagala M and Lewandowski J 2004 \CQG \textbf{21} 5233; Engle J, Noui K and Perez A 2010 \PRL \textbf{105}  031302.
\bibitem{lt} Livine E R and Terno D R 2008 \NP B \textbf{794} 138; Livine E R and Terno D R 2009 \NP B \textbf{806} 715.
\bibitem{ash03} Ashtekar A, Fairhurst S and Willis J L 2003 \CQG
\textbf{20} 1031.
\bibitem{gs-ln} Giesel K and Sahlman H 2012 {\it Preprint} arXiv:1203.2733.
\bibitem{sigma} Chacon-Acosta G, Manrique E, Dagdud L and Morales-TŽcotl H A 2011 {\it SIGMA} \textbf{7} 110.
\bibitem{halv} Halvorson H 2004 {\it Studies Hist. Philos. Mod. Phys.} \textbf{35} 45.
\bibitem{gam69} Gamelin T 1969 {\it Uniform Algebras}  (Englewood Cliffs NJ: Prentice-Hall).
\bibitem{fred06} Fredenhagen F and Reszewski F 2006 \CQG \textbf{23} 6577.
\bibitem{cor07} Corichi A, Vuka\v{s}inac T and Zapata J A 2007 \PR D \textbf{76} 044016; Corichi A, Vuka\v{s}inac T and Zapata J A 2007 \CQG \textbf{24} 1495.
{\bibitem{velhinho} Velhinho J M 2007 \CQG \textbf{24} 3745.}
\bibitem{br2} Bratteli O and Robinson D W 1992 {\it Operator Algebras
and Quantum Statistical Mechanics} vol.~2 (Berlin: Springer).
\bibitem{bohr_wz} Bohr N 1949 in {\it Albert Einstein: Philosopher-Scientist} ed
  P A Schilpp (Evanston IL: Library of Living Philosophers) p 199; Ionicioiu R and Terno D R 2011 \PRL {\bf 107} 230406.
\bibitem{app} Husain V, Louko J and Winkler O 2007 \PR D \textbf{76} 084002; Kunstatter G and Louko J 2012 {\it Preprint} arXiv:1201.2886.
\bibitem{viqar} Husain V and Winkler O 2005 \CQG \textbf{22}
L127; Husain V and Winkler O 2005 \CQG \textbf{22} L135; Husain V and Terno D R 2010 \PR D \textbf{81}  044039.
\bibitem{fell} Fell J M G 1960 {\it Proc. Am. Math. Soc.} \textbf{94} 365; Wald R M 1994 {\it Quantum Field Theory in
Curved Spacetime and Black Hole Thermodynamics} (Chicago: University of Chicago Press).
\bibitem{gauss} Braunstein S and van Look P 2005 \RMP \textbf{77} 513; Wang X-B, Hiroshima T, Tomita A and Hayashi M 2007 {\it Phys. Rep.} \textbf{448} 1; Olivares S 2012 {\it Eur. Phys. J. Special Topics} \textbf{203} 3.
\bibitem{tom} Demarie T F 2012 {\it Preprint} arXiv:1209.2748.
\bibitem{gqi} Weedbrook C et al. 2012 \RMP \textbf{84} 621.
\bibitem{fabio} Jizba P, Kleinert H and Scardigli F 2010 \PR D \textbf{81} 084030; Husain V, Kothawala D and Seahra S S 2012 {\it Preprint} arXiv:1208.5761.


\end{thebibliography}
\end{document}